\documentclass[a4paper,aps,prd,10pt,preprintnumbers,showpacs,twocolumn,superscriptaddress,nofootinbib,amsmath,amssymb]{revtex4-1}
\usepackage{graphicx}
\usepackage{cmap}
\usepackage[utf8]{inputenc}
\usepackage[T1]{fontenc}

\begin{document}
\title{Long-lived quasinormal modes and grey-body factors of supermassive black holes with a dark matter halo}
\author{Zainab Malik}\email{zainabmalik8115@outlook.com}
\affiliation{Institute of Applied Sciences and Intelligent Systems, H-15, Pakistan}

\begin{abstract}
We study quasinormal modes and grey-body factors of a massive scalar field in the background of a Schwarzschild black hole surrounded by a spherically symmetric galactic dark matter halo.  
The background metric, recently obtained as an analytic generalization of the Schwarzschild geometry, depends on the halo velocity parameter $V_{c}$ and the core radius $a$.  
Using the sixth- and seventh-order WKB methods with Padé approximants, supported by time-domain integration and Prony analysis, we compute the fundamental quasinormal frequencies and transmission coefficients. The results show that the real part of the frequency slightly increases while the damping rate decreases with growing field mass $\mu$, leading to longer-lived oscillations.  
The influence of the dark matter halo parameters is found to be negligible for astrophysically realistic values, confirming the robustness of Schwarzschild-like ringdown signatures.  
Grey-body factors decrease with increasing field mass and multipole number, while the effect of the halo parameters remains small. 
\end{abstract}

\maketitle

\section{Introduction}

The analysis of quasinormal modes (QNMs) of black holes \cite{Kokkotas:1999bd,Nollert:1999ji,Bolokhov:2025uxz,Konoplya:2011qq} provides a direct probe of how external environments influence the dynamical response of compact objects \cite{Kokkotas:1999bd,Nollert:1999ji,Bolokhov:2025uxz}. In astrophysical settings, black holes are rarely isolated. Observational and theoretical evidence suggests that supermassive and stellar-mass black holes are embedded in galactic halos composed predominantly of dark matter. Although the local dark-matter density near the black-hole horizon is small compared to the mean energy density of the hole itself, even a weakly interacting dark component can modify the background geometry or act as an external potential affecting the propagation of perturbative fields. Therefore, quasinormal modes and other characteristics of radiation phenomena around such black holes immersed in galactic halo have been intensively studied recently \cite{Liu:2024xcd,Chen:2024lpd,Hamil:2025pte,Konoplya:2025nqv,Pezzella:2024tkf,Becar:2024agj,Liu:2023vno,Jha:2024ltc,Liu:2024bfj,Mollicone:2024lxy,Dubinsky:2025fwv,Konoplya:2025ect,Malik:2026lfj,Bolokhov:2026eqf,Bolokhov:2025zva,Bolokhov:2025fto,Saka:2026ott,Lutfuoglu:2026fks,Lutfuoglu:2026zel,Lutfuoglu:2025mqa}.

A practical and physically transparent way to model the influence of the dark-matter halo is to consider the propagation of a test scalar field with an effective mass term $\mu_{\rm eff}$ in the fixed black-hole background. The field equation then reads
\begin{equation}
\Box \Phi - \mu_{\rm eff}^{2} \Phi = 0,
\end{equation}
where the effective mass $\mu_{\rm eff}$ may incorporate several physical effects. 
When $\mu_{\rm eff}$ originates from a self-interacting scalar or a dark-sector particle species, it parametrizes the local interaction strength with the surrounding halo. Alternatively, $\mu_{\rm eff}$ can mimic couplings to other background fields, such as an external magnetic field or a plasma frequency, which likewise generate an effective potential term proportional to the square of the field amplitude \cite{Konoplya:2007yy,Kokkotas:2010zd,Konoplya:2008hj,Chen:2011jgd,Davlataliev:2024mjl}. In this sense, a ``massive'' scalar field provides a universal phenomenological framework to study a broad class of environmental influences on black-hole perturbations.

The introduction of an effective mass term modifies both the real and imaginary parts of the QNM spectrum. The effective potential barrier governing wave propagation acquires a nonvanishing asymptotic limit, leading to the appearance of quasibound or long-lived modes, often referred to as \emph{quasi-resonances} \cite{Ohashi:2004wr,Konoplya:2017tvu}. These modes can significantly alter the late-time relaxation of perturbations and produce distinct signatures in the gravitational-wave ringdown spectrum \cite{Jing:2004zb,Koyama:2001qw,Moderski:2001tk,Rogatko:2007zz,Koyama:2001ee,Churilova:2019qph,Dubinsky:2024jqi}. From the observational point of view, understanding how such environmental terms modify the QNM spectrum is crucial for interpreting deviations from the Kerr paradigm in current and future gravitational-wave detections. The massive fields may also contribute to the very long wave radiation observed via the Pulsar Timing Array \cite{Konoplya:2023fmh}. After all, an effective mass term may appear as the tidal effect from the bulk in the brane-world models \cite{Seahra:2004fg,Ishihara:2008re}. 

Moreover, the study of QNMs in dark-matter–inspired spacetimes is closely related to the problem of stability and energy exchange between the black hole and its surroundings. The presence of a massive field can give rise to superradiant instabilities when the effective potential allows partial wave trapping. Therefore, exploring how the parameters of the dark-matter halo or of the effective mass influence the spectrum of quasinormal frequencies provides insight into the dynamical stability of realistic astrophysical black holes and the potential observational imprints of dark matter or external fields in their vicinity. 

Having all of the above motivations in mind, perturbations, quasinormal modes and scattering of massive fields have been extensively studied in \cite{Konoplya:2007zx,Konoplya:2024wds,Fernandes:2021qvr,Percival:2020skc,Bolokhov:2023ruj,Bolokhov:2023bwm,Bolokhov:2024bke,Bolokhov:2024ixe,Skvortsova:2025cah,Skvortsova:2024eqi,Lutfuoglu:2025qkt,Lutfuoglu:2025bsf,Lutfuoglu:2025hwh,Lutfuoglu:2025hjy,Dubinsky:2024hmn,Zinhailo:2024jzt,Zinhailo:2024kbq,Malik:2025ava}.

For these reasons, quasinormal-mode analysis of black holes immersed in dark-matter environments has become a natural extension of classical black-hole perturbation theory. It bridges astrophysical modeling with fundamental physics by linking gravitational-wave observables to properties of the surrounding matter distribution and possible new degrees of freedom beyond the Standard Model. For the black hole in galactic halo constructed in \\\cite{Lobo:2025kzb}  the quasinormal modes of massless fields have been recently considered in \cite{Dubinsky:2025fwv}.

Another chracteristic we are interested in this pape is grey-body factors of a massive scalar field. Grey-body factors and Hawking emission of various black holes have been considered in  \cite{Page:1976df,Page:1976ki,Kanti:2002nr}.  
In the context of gauge/gravity duality, grey-body factors govern the poles of the transmission coefficient correspond to those of the retarded Green’s functions, linking black-hole scattering to the optical and electromagnetic response of holographic plasmas and superconductors ~\cite{Horowitz:2008bn,Herzog:2009xv,Konoplya:2009hv}.

Having in mind the above motivations,  we will study quasinormal modes and grey-body factors of a massive scalar field in the background of the black holes immersed in a dark matter halo constructed in \cite{Lobo:2025kzb,Lobo:2025qap}.

The paper is organized as follows.  
In Sec.~II, we review the analytic form of the Schwarzschild black hole immersed in a galactic dark matter halo and summarize its basic geometric properties.  
Section~III outlines the methods employed for determining the quasinormal spectrum, including the higher-order WKB approach with Padé approximants and the time-domain integration technique with Prony analysis.  
In Sec.~IV, we present and discuss the quasinormal frequencies of the massive scalar field and analyze their dependence on the field mass and halo parameters.  
Section~V is devoted to the computation of grey-body factors, where we study their variation with frequency, angular momentum, and the dark halo characteristics.  
Finally, Sec.~VI summarizes the main results and conclusions.

\section{Black Hole immersed in the Galactic Halo}\label{sec:background}

We consider a static, spherically symmetric black hole embedded in a galactic dark-matter halo, whose spacetime geometry was recently obtained as an analytic generalization of the Schwarzschild metric~\cite{Lobo:2025kzb,Ma:2024oqe,Lobo:2025qap}. 
The halo is characterized by a physically motivated density distribution that reproduces the nearly flat galactic rotation curves~\cite{Shen:2009my}:
\begin{equation}
\rho(r)=\frac{V_c^2}{4\pi G}\,\frac{3a^2+r^2}{(a^2+r^2)^2},
\label{density-profile}
\end{equation}
where the parameters $V_c$ and $a$ denote, respectively, the asymptotic circular velocity of the galaxy and the core radius of the dark-matter distribution.

To determine the corresponding spacetime metric, we adopt the general static and spherically symmetric line element
\begin{equation}
ds^2=-\bigl[1+F(r)\bigr]\,dt^2+\bigl[1+F(r)\bigr]^{-1}dr^2+r^2d\Omega^2,
\label{ansatz}
\end{equation}
and solve the Einstein equations for $F(r)$ with the energy density profile~\eqref{density-profile} acting as the matter source.  
Integration of the field equations yields
\begin{equation}
F(r)=\frac{C_1}{r}-\frac{2V_c^2r^2}{a^2+r^2},
\end{equation}
where $C_1$ is an integration constant. Notice that in ~\cite{Lobo:2025kzb} the sign in front of the last term is wrong.
Requiring the metric to approach the Schwarzschild form when the halo contribution vanishes ($V_c\to0$) fixes this constant to $C_1=-2M$, with $M$ interpreted as the black-hole mass.

The full metric can then be expressed in the compact form
\begin{equation}
ds^2=-f(r)\,dt^2+f^{-1}(r)\,dr^2+r^2d\Omega^2,
\label{metric-halo}
\end{equation}
with the lapse function
\begin{equation}
f(r)=1-\frac{2M}{r}-\frac{2V_c^2r^2}{a^2+r^2}.
\end{equation}
This solution represents a Schwarzschild black hole surrounded by a spherically symmetric galactic halo.  
In the absence of the halo contribution ($V_c=0$), it smoothly reduces to the vacuum Schwarzschild geometry.

\begin{figure}
\resizebox{\linewidth}{!}{\includegraphics{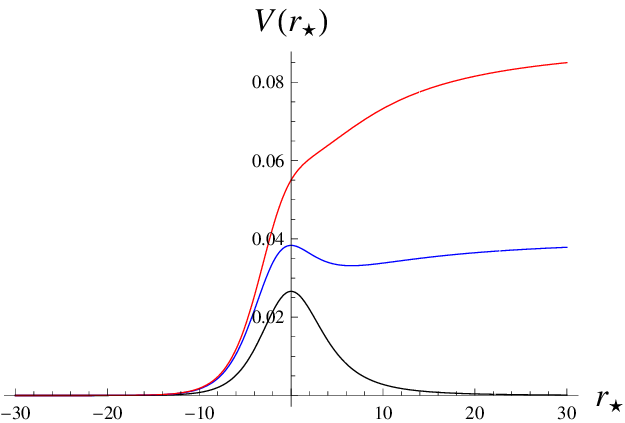}}
\caption{Effective potentials for $\ell=0$ massive scalar field perturbations at $M=1$, $V_{c}=0.1$, $a=10$, $\mu=0$ (black), $\mu=0.2$ (blue) and $\mu=0.3$ (red).}\label{fig:potential1}
\end{figure}

\begin{figure}
\resizebox{\linewidth}{!}{\includegraphics{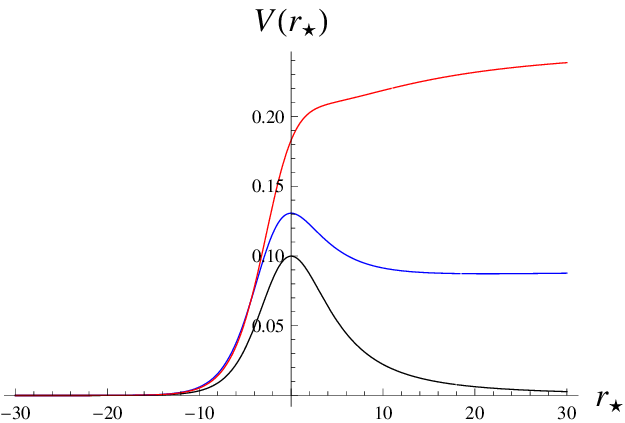}}
\caption{Effective potentials for $\ell=1$ massive scalar field perturbations at $M=1$, $V_{c}=0.1$, $a=10$, $\mu=0$ (black), $\mu=0.3$ (blue) and $\mu=0.5$ (red).}\label{fig:potential2}
\end{figure}

The event horizon $r_h$ is determined as the largest positive root of $f(r)=0$, that is,
\begin{equation}
1-\frac{2M}{r}-\frac{2V_c^2r^2}{a^2+r^2}=0.
\end{equation}
Expanding this condition perturbatively for small $V_c$ gives
\begin{equation}\nonumber
r_h
=2M+\frac{16M^3}{a^2+4M^2} V_c^2
+\frac{128M^5(3a^2+4M^2)}{(a^2+4M^2)^3} V_c^4
+\mathcal{O}(V_c^6).
\end{equation}

In what follows, we use the analyticity of the metric~\eqref{metric-halo} to investigate quasinormal spectra of a massive scalar perturbations—and analyze how the dark-matter halo parameters affect their dynamical behavior.

\begin{table}
\begin{tabular}{c l c c l}
\hline
$\ell$ & $\mu$  & WKB-6, $m=3$ & WKB-7, $m=3$ &  $\delta (\%)$ \\
\hline
$0$ & $0$ & $0.111682-0.104740 i$ & $0.111682-0.104748 i$ & $0.0054\%$\\
$0$ & $0.01$ & $0.111706-0.104659 i$ & $0.111706-0.104661 i$ & $0.0016\%$\\
$0$ & $0.05$ & $0.112249-0.102713 i$ & $0.112228-0.102585 i$ & $0.0855\%$\\
$0$ & $0.1$ & $0.113638-0.096752 i$ & $0.113482-0.096574 i$ & $0.159\%$\\
$0$ & $0.15$ & $0.114914-0.087411 i$ & $0.114557-0.086609 i$ & $0.608\%$\\
%$0$ & $0.2$ & $0.117454-0.091920 i$ & $0.114702-0.090519 i$ & $2.07\%$\\
%$0$ & $0.25$ & $0.214013-0.000096 i$ & $0.218567-0.037738 i$ & $17.7\%$\\
$1$ & $0$ & $0.292036-0.097467 i$ & $0.292038-0.097467 i$ & $0.0006\%$\\
$1$ & $0.01$ & $0.292081-0.097440 i$ & $0.292082-0.097440 i$ & $0.0006\%$\\
$1$ & $0.05$ & $0.293154-0.096798 i$ & $0.293155-0.096797 i$ & $0.0005\%$\\
$1$ & $0.1$ & $0.296515-0.094775 i$ & $0.296516-0.094775 i$ & $0.0003\%$\\
$1$ & $0.15$ & $0.302141-0.091353 i$ & $0.302139-0.091353 i$ & $0.0004\%$\\
$1$ & $0.2$ & $0.310067-0.086445 i$ & $0.310058-0.086446 i$ & $0.0028\%$\\
$1$ & $0.25$ & $0.320339-0.079917 i$ & $0.320306-0.079915 i$ & $0.0102\%$\\
$1$ & $0.3$ & $0.332986-0.071626 i$ & $0.332884-0.071529 i$ & $0.0413\%$\\
$1$ & $0.35$ & $0.347812-0.061196 i$ & $0.347669-0.061210 i$ & $0.0405\%$\\
$1$ & $0.4$ & $0.364602-0.048490 i$ & $0.364729-0.048295 i$ & $0.0633\%$\\
%$1$ & $0.45$ & $0.397921-0.030269 i$ & $0.376275-0.035665 i$ & $5.59\%$\\
$2$ & $0$ & $0.482338-0.096516 i$ & $0.482338-0.096515 i$ & $0.00014\%$\\
$2$ & $0.01$ & $0.482369-0.096505 i$ & $0.482370-0.096504 i$ & $0.00014\%$\\
$2$ & $0.05$ & $0.483126-0.096247 i$ & $0.483127-0.096246 i$ & $0.00014\%$\\
$2$ & $0.1$ & $0.485493-0.095437 i$ & $0.485494-0.095437 i$ & $0.00013\%$\\
$2$ & $0.15$ & $0.489448-0.094082 i$ & $0.489448-0.094081 i$ & $0.00013\%$\\
$2$ & $0.2$ & $0.495003-0.092170 i$ & $0.495003-0.092170 i$ & $0.00011\%$\\
$2$ & $0.25$ & $0.502179-0.089686 i$ & $0.502179-0.089686 i$ & $0.00007\%$\\
$2$ & $0.3$ & $0.511001-0.086609 i$ & $0.511001-0.086609 i$ & $0\%$\\
$2$ & $0.35$ & $0.521504-0.082909 i$ & $0.521504-0.082910 i$ & $0.00013\%$\\
$2$ & $0.4$ & $0.533729-0.078546 i$ & $0.533727-0.078546 i$ & $0.00041\%$\\
$2$ & $0.45$ & $0.547727-0.073465 i$ & $0.547720-0.073464 i$ & $0.00119\%$\\
$2$ & $0.5$ & $0.563545-0.067595 i$ & $0.563539-0.067584 i$ & $0.00230\%$\\
$2$ & $0.55$ & $0.581246-0.060810 i$ & $0.581238-0.060799 i$ & $0.00235\%$\\
$2$ & $0.6$ & $0.600885-0.052963 i$ & $0.600876-0.052959 i$ & $0.00153\%$\\
$2$ & $0.65$ & $0.622430-0.043869 i$ & $0.622460-0.043845 i$ & $0.00610\%$\\
$2$ & $0.7$ & $0.646529-0.034289 i$ & $0.645744-0.032949 i$ & $0.240\%$\\
\hline
\end{tabular}
\caption{Quasinormal modes of the black hole immersed in galactic halo ($M=1$, $V_{c}=0.1$, $a=10$) calculated using the WKB formula at various orders.}
\end{table}

\begin{table}
\begin{tabular}{c l c c l}
\hline
$\ell$ & $\mu$  & WKB-6, $m=3$ & WKB-7, $m=3$ &  $\delta (\%)$ \\
\hline
$0$ & $0$ & $0.110632-0.104980 i$ & $0.110619-0.105668 i$ & $0.452\%$\\
$0$ & $0.01$ & $0.110660-0.104900 i$ & $0.110649-0.105584 i$ & $0.449\%$\\
$0$ & $0.05$ & $0.111294-0.102977 i$ & $0.111354-0.103548 i$ & $0.379\%$\\
$0$ & $0.1$ & $0.112834-0.096988 i$ & $0.113059-0.097222 i$ & $0.219\%$\\
$0$ & $0.15$ & $0.115132-0.087667 i$ & $0.115292-0.086896 i$ & $0.545\%$\\
$0$ & $0.2$ & $0.113147-0.088370 i$ & $0.114749-0.089334 i$ & $1.30\%$\\
%$0$ & $0.25$ & $0.202440-0.002928 i$ & $0.218534-0.072299 i$ & $35.2\%$\\
$1$ & $0$ & $0.289343-0.096876 i$ & $0.289342-0.096876 i$ & $0.0004\%$\\
$1$ & $0.01$ & $0.289387-0.096850 i$ & $0.289386-0.096850 i$ & $0.0004\%$\\
$1$ & $0.05$ & $0.290459-0.096215 i$ & $0.290458-0.096216 i$ & $0.0005\%$\\
$1$ & $0.1$ & $0.293817-0.094219 i$ & $0.293815-0.094220 i$ & $0.0008\%$\\
$1$ & $0.15$ & $0.299441-0.090841 i$ & $0.299435-0.090842 i$ & $0.0018\%$\\
$1$ & $0.2$ & $0.307371-0.085995 i$ & $0.307357-0.085999 i$ & $0.0045\%$\\
$1$ & $0.25$ & $0.317665-0.079543 i$ & $0.317626-0.079545 i$ & $0.0118\%$\\
$1$ & $0.3$ & $0.330387-0.071293 i$ & $0.330273-0.071234 i$ & $0.0380\%$\\
$1$ & $0.35$ & $0.345397-0.060927 i$ & $0.344946-0.060593 i$ & $0.160\%$\\
$1$ & $0.4$ & $0.362375-0.047804 i$ & $0.362221-0.048107 i$ & $0.0928\%$\\
$1$ & $0.45$ & $0.382708-0.034927 i$ & $0.380035-0.031710 i$ & $1.09\%$\\
$2$ & $0$ & $0.478407-0.095780 i$ & $0.478408-0.095780 i$ & $0.00012\%$\\
$2$ & $0.01$ & $0.478438-0.095770 i$ & $0.478439-0.095769 i$ & $0.00012\%$\\
$2$ & $0.05$ & $0.479192-0.095515 i$ & $0.479192-0.095515 i$ & $0.00012\%$\\
$2$ & $0.1$ & $0.481549-0.094719 i$ & $0.481549-0.094719 i$ & $0.00011\%$\\
$2$ & $0.15$ & $0.485486-0.093386 i$ & $0.485486-0.093386 i$ & $0.00010\%$\\
$2$ & $0.2$ & $0.491017-0.091507 i$ & $0.491018-0.091507 i$ & $0.00008\%$\\
$2$ & $0.25$ & $0.498163-0.089067 i$ & $0.498163-0.089067 i$ & $0.00005\%$\\
$2$ & $0.3$ & $0.506949-0.086048 i$ & $0.506949-0.086048 i$ & $0.00002\%$\\
$2$ & $0.35$ & $0.517410-0.082423 i$ & $0.517410-0.082423 i$ & $0.00014\%$\\
$2$ & $0.4$ & $0.529590-0.078154 i$ & $0.529588-0.078155 i$ & $0.00041\%$\\
$2$ & $0.45$ & $0.543545-0.073194 i$ & $0.543539-0.073193 i$ & $0.00113\%$\\
$2$ & $0.5$ & $0.559332-0.067477 i$ & $0.559326-0.067464 i$ & $0.00241\%$\\
$2$ & $0.55$ & $0.577025-0.060882 i$ & $0.577025-0.060862 i$ & $0.00359\%$\\
$2$ & $0.6$ & $0.596721-0.053256 i$ & $0.596705-0.053217 i$ & $0.00700\%$\\
$2$ & $0.65$ & $0.618479-0.044331 i$ & $0.618414-0.044296 i$ & $0.0120\%$\\
$2$ & $0.7$ & $0.642141-0.033729 i$ & $0.642191-0.033746 i$ & $0.00824\%$\\
$2$ & $0.75$ & $0.673496-0.022348 i$ & $0.665928-0.021761 i$ & $1.13\%$\\
\hline
\end{tabular}
\caption{Quasinormal modes of the black hole immersed in galactic halo ($M=1$, $V_{c}=0.2$, $a=10$) calculated using the WKB formula at various orders.}
\end{table}

\begin{table}
\begin{tabular}{c l c c l}
\hline
$\ell$ & $\mu$  & WKB-6, $m=3$ & WKB-7, $m=3$ &  $\delta (\%)$ \\
\hline
$0$ & $0$ & $0.100566-0.105652 i$ & $0.100132-0.107858 i$ & $1.54\%$\\
$0$ & $0.01$ & $0.100602-0.105578 i$ & $0.100169-0.107789 i$ & $1.55\%$\\
$0$ & $0.05$ & $0.101442-0.103802 i$ & $0.101038-0.106149 i$ & $1.64\%$\\
$0$ & $0.1$ & $0.103730-0.098255 i$ & $0.103582-0.101156 i$ & $2.03\%$\\
$0$ & $0.15$ & $0.105957-0.089515 i$ & $0.107161-0.093944 i$ & $3.31\%$\\
$0$ & $0.2$ & $0.107353-0.084326 i$ & $0.109201-0.086483 i$ & $2.08\%$\\
%$0$ & $0.25$ & $0.147119-0.063382 i$ & $0.096113-0.086081 i$ & $34.9\%$\\
%$0$ & $0.3$ & $0.217608-0.009911 i$ & $0.218265-0.021029 i$ & $5.11\%$\\
$0$ & $0.35$ & $0.256370-0.014990 i$ & $0.255228-0.015850 i$ & $0.557\%$\\
$0$ & $0.4$ & $0.294215-0.015995 i$ & $0.293461-0.016524 i$ & $0.313\%$\\
$0$ & $0.45$ & $0.331363-0.016157 i$ & $0.331216-0.016981 i$ & $0.252\%$\\
$0$ & $0.5$ & $0.368701-0.017063 i$ & $0.368843-0.017273 i$ & $0.0688\%$\\
$0$ & $0.55$ & $0.406350-0.017412 i$ & $0.406404-0.017472 i$ & $0.0198\%$\\
$0$ & $0.6$ & $0.443897-0.017596 i$ & $0.443918-0.017617 i$ & $0.00662\%$\\
$0$ & $0.65$ & $0.481386-0.017720 i$ & $0.481396-0.017726 i$ & $0.00239\%$\\
$0$ & $0.7$ & $0.518838-0.017811 i$ & $0.518843-0.017811 i$ & $0.00095\%$\\
$0$ & $0.75$ & $0.556264-0.017880 i$ & $0.556265-0.017877 i$ & $0.00049\%$\\
$1$ & $0$ & $0.270166-0.092285 i$ & $0.270149-0.092295 i$ & $0.0067\%$\\
$1$ & $0.01$ & $0.270210-0.092261 i$ & $0.270193-0.092270 i$ & $0.0067\%$\\
$1$ & $0.05$ & $0.271268-0.091683 i$ & $0.271251-0.091692 i$ & $0.0069\%$\\
$1$ & $0.1$ & $0.274588-0.089864 i$ & $0.274568-0.089873 i$ & $0.0077\%$\\
$1$ & $0.15$ & $0.280168-0.086793 i$ & $0.280143-0.086802 i$ & $0.0091\%$\\
$1$ & $0.2$ & $0.288088-0.082403 i$ & $0.288055-0.082408 i$ & $0.0112\%$\\
$1$ & $0.25$ & $0.298506-0.076581 i$ & $0.298465-0.076572 i$ & $0.0139\%$\\
$1$ & $0.3$ & $0.311743-0.069187 i$ & $0.311708-0.069128 i$ & $0.0216\%$\\
$1$ & $0.35$ & $0.328341-0.060128 i$ & $0.328499-0.060022 i$ & $0.0572\%$\\
$1$ & $0.4$ & $0.349617-0.049923 i$ & $0.349866-0.050056 i$ & $0.0800\%$\\
$1$ & $0.45$ & $0.376094-0.040821 i$ & $0.376166-0.040815 i$ & $0.0192\%$\\
$1$ & $0.5$ & $0.406054-0.033770 i$ & $0.406219-0.033777 i$ & $0.0406\%$\\
$1$ & $0.55$ & $0.437936-0.028985 i$ & $0.438379-0.029036 i$ & $0.101\%$\\
$1$ & $0.6$ & $0.471025-0.025803 i$ & $0.471650-0.025763 i$ & $0.133\%$\\
$1$ & $0.65$ & $0.504625-0.022851 i$ & $0.505745-0.023370 i$ & $0.244\%$\\
$1$ & $0.7$ & $0.539290-0.021614 i$ & $0.539503-0.022124 i$ & $0.102\%$\\
$1$ & $0.75$ & $0.574553-0.020983 i$ & $0.574396-0.021123 i$ & $0.0365\%$\\
$2$ & $0$ & $0.450221-0.090400 i$ & $0.450221-0.090400 i$ & $0\%$\\
$2$ & $0.01$ & $0.450252-0.090391 i$ & $0.450252-0.090391 i$ & $0\%$\\
$2$ & $0.05$ & $0.450980-0.090165 i$ & $0.450980-0.090165 i$ & $0\%$\\
$2$ & $0.1$ & $0.453259-0.089460 i$ & $0.453259-0.089460 i$ & $0\%$\\
$2$ & $0.15$ & $0.457065-0.088283 i$ & $0.457065-0.088283 i$ & $0\%$\\
$2$ & $0.2$ & $0.462412-0.086631 i$ & $0.462412-0.086631 i$ & $0\%$\\
$2$ & $0.25$ & $0.469319-0.084501 i$ & $0.469319-0.084501 i$ & $0\%$\\
$2$ & $0.3$ & $0.477814-0.081888 i$ & $0.477814-0.081888 i$ & $0\%$\\
$2$ & $0.35$ & $0.487932-0.078787 i$ & $0.487932-0.078787 i$ & $0\%$\\
$2$ & $0.4$ & $0.499721-0.075194 i$ & $0.499721-0.075194 i$ & $0.00002\%$\\
$2$ & $0.45$ & $0.513247-0.071111 i$ & $0.513246-0.071109 i$ & $0.00049\%$\\
$2$ & $0.5$ & $0.528589-0.066537 i$ & $0.528591-0.066538 i$ & $0.00040\%$\\
$2$ & $0.55$ & $0.545874-0.061511 i$ & $0.545876-0.061513 i$ & $0.00046\%$\\
$2$ & $0.6$ & $0.565237-0.056114 i$ & $0.565253-0.056127 i$ & $0.00357\%$\\
$2$ & $0.65$ & $0.586840-0.050683 i$ & $0.586855-0.050547 i$ & $0.0234\%$\\
$2$ & $0.7$ & $0.610755-0.045127 i$ & $0.610757-0.045064 i$ & $0.0103\%$\\
$2$ & $0.75$ & $0.636859-0.040054 i$ & $0.636861-0.040036 i$ & $0.00290\%$\\
\hline
\end{tabular}
\caption{Quasinormal modes of the black hole immersed in galactic halo ($M=1$, $V_{c}=0.5$, $a=10$) calculated using the WKB formula at various orders.}
\end{table}

\begin{figure}
\resizebox{\linewidth}{!}{\includegraphics{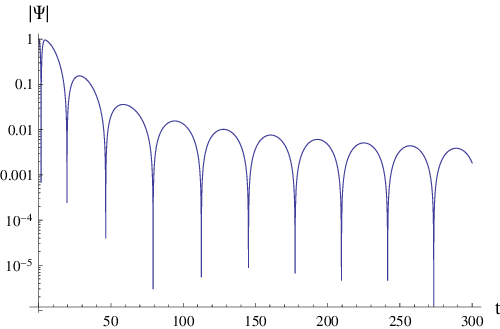}}
\caption{Semi-logarithmic time-domain profile for $\ell=0$ perturbations. Here $\mu=0.1$, $V_{c}=0.1$, $a=10$. The WKB data is $0.113638-0.096752 i$, and the extraction the dominant frequency from the time-domain profile via the Prony method gives $\omega = 0.121193 - 0.0955343 i$. This does not mean immediately insufficient accuracy of the WKB data, because the period of quasinormal oscillations is quickly changed by the late-time tail at $t \approx 80$.}\label{fig:TDL0}
\end{figure}

\begin{figure}
\resizebox{\linewidth}{!}{\includegraphics{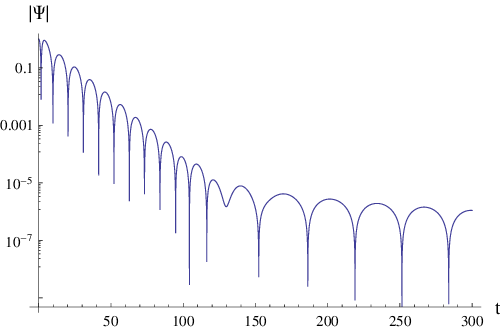}}
\caption{Semi-logarithmic time-domain profile for $\ell=1$ perturbations. Here $\mu=0.1$, $V_{c}=0.1$, $a=10$. The WKB data is $0.296515 - 0.094775 i$, and the extraction the dominant frequency from the time-domain profile via the Prony method gives $\omega = 0.296517 - 0.0947754 i$.}\label{fig:TDL1}
\end{figure}

\section{Methods for finding quasinormal modes}

Perturbations of a black hole by external fields produce characteristic damped oscillations known as \emph{quasinormal modes} (QNMs).  
They describe how the spacetime reacts to small disturbances and return to equilibrium through the emission of radiation.  
Formally, QNMs correspond to the solutions of the linearized perturbation equations satisfying boundary conditions appropriate to energy loss through the horizon and to spatial infinity.

For a static, spherically symmetric black hole, the scalar field can be separated into angular and radial parts,
\begin{equation}
\Phi(t,r,\theta,\phi)
=\sum_{\ell,m}\frac{\Psi_{\ell m}(r)}{r}\,
Y_{\ell m}(\theta,\phi)\,e^{-i\omega t},
\end{equation}
leading to the Schrödinger-type master equation
\begin{equation}
\frac{d^2\Psi_{\ell}}{dr_*^2}
+\bigl[\omega^2 - V_{\ell}(r)\bigr]\Psi_{\ell}=0,
\label{wave_eq}
\end{equation}
where the tortoise coordinate is defined by \(dr_*/dr=1/f(r)\), and the effective potential for a minimally coupled scalar field of mass \(\mu\) reads
\begin{equation}
V_{\ell}(r)=
f(r)\!\left[\frac{\ell(\ell+1)}{r^2}+\frac{f'(r)}{r}\right]
+f(r)\,\mu^2.
\label{potential}
\end{equation}

Unlike the massless case, the potential~\eqref{potential} approaches a nonzero constant,
\(\displaystyle \lim_{r\to\infty}V_{\ell}(r)=\mu^2\), so the asymptotic form of the wave function depends on the relative magnitude of \(\omega\) and \(\mu\).  
The effective potentials are shown in figs. \ref{fig:potential1} and \ref{fig:potential2}, where one can see that at sufficiently large mass of the field, the peak disappears.

In terms of the tortoise coordinate, the boundary conditions are:
\begin{widetext}
\begin{equation}
\Psi_{\ell}(r_*) \sim
\begin{cases}
e^{-i\omega r_*}, & r_* \rightarrow -\infty 
\quad (\text{purely ingoing at the horizon}),\\[6pt]
e^{+i\sqrt{\omega^2-\mu^2}\,r_*}, & r_* \rightarrow +\infty, \quad (\text{outgoing wave}),\\[6pt]
\end{cases}
\label{boundary_massive}
\end{equation}
\end{widetext}
The resulting spectrum consists of complex frequencies
\(\omega_{\ell n}=\omega_{\mathrm{R}}-i\omega_{\mathrm{I}}\),   where \(\omega_{\mathrm{R}}\) determines the oscillation frequency and $\omega_{\mathrm{I}}$ is proportional to the damping rate. 

\begin{figure}
\resizebox{\linewidth}{!}{\includegraphics{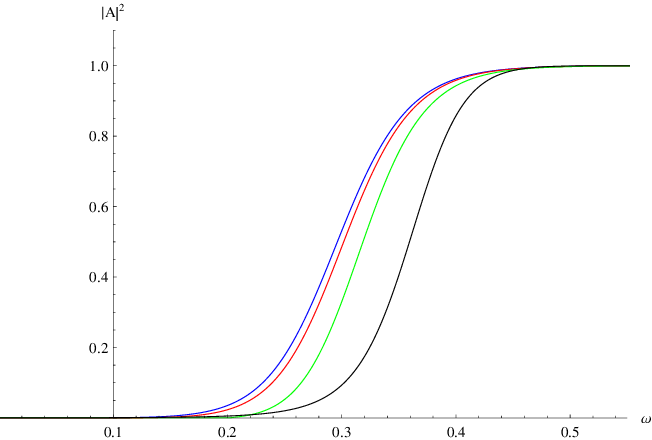}}
\caption{Grey-body factors for $\ell=1$ massive scalar field perturbations at $M=1$, $V_{c}=0.1$, $a=10$, $\mu=0$ (blue), $\mu=0.1$ (red), $\mu=0.2$ (green), and $\mu=0.3$ (black).}\label{fig:GBFl1}
\end{figure}

\begin{figure}
\resizebox{\linewidth}{!}{\includegraphics{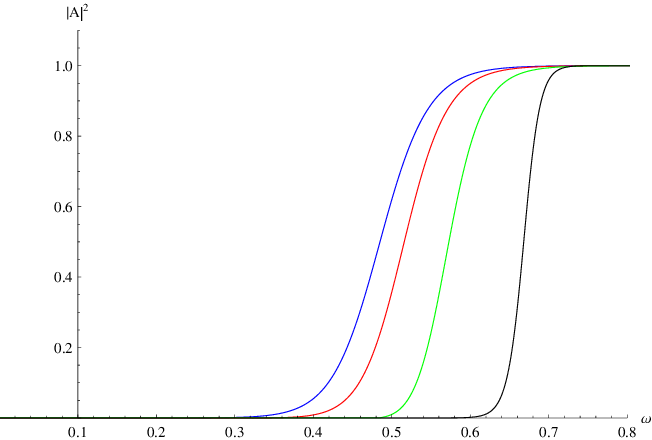}}
\caption{Grey-body factors for $\ell=2$ massive scalar field perturbations at $M=1$, $V_{c}=0.1$, $a=10$, $\mu=0$ (blue), $\mu=0.3$ (red), $\mu=0.5$ (green), and $\mu=0.7$ (black).}\label{fig:GBFl2}
\end{figure}

\begin{figure}
\resizebox{\linewidth}{!}{\includegraphics{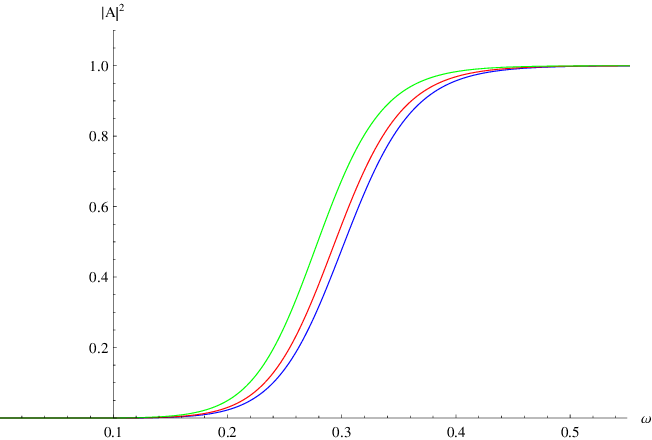}}
\caption{Grey-body factors for $\ell=1$ massive scalar field perturbations at $M=1$, $\mu=0.1$, $a=10$, $V_{c}=0$ (blue), $V_{c}=0.3$ (red), $V_{c}=0.5$ (green).}\label{fig:GBFl1Vc}
\end{figure}

\subsection{WKB method with Padé approximants}

To compute the quasinormal spectrum and grey-body factors in the background~\eqref{metric-halo}, we employ the higher-order WKB approximation supplemented by Padé approximants.  
The WKB method provides an accurate semi-analytic description of wave scattering in single-barrier potentials of the type that arise from black-hole perturbations~\cite{Konoplya:2001ji,Kodama:2009bf,Bolokhov:2025egl,Skvortsova:2024wly,Skvortsova:2024atk,Skvortsova:2023zmj,Bolokhov:2022rqv,Lutfuoglu:2025pzi,Lutfuoglu:2025blw,Lutfuoglu:2025ljm,Momennia:2022tug,Zhao:2022gxl,del-Corral:2022kbk}.

In the vicinity of the potential peak $r=r_0$, the WKB quantization condition for complex frequencies $\omega$ takes the form
\begin{equation}
i\,\frac{\omega^2 - V_0}{\sqrt{-2V_0''}}
- \sum_{i=2}^{N} \Lambda_i = n+\frac{1}{2},
\label{wkb-condition}
\end{equation}
where $V_0$ and $V_0''$ denote the value and the second derivative of the effective potential at its maximum, $n$ is the overtone number, and the $\Lambda_i$ are higher-order WKB corrections explicitly given in~\cite{Schutz:1985km,Iyer:1986np,Konoplya:2003ii,Matyjasek:2017psv}.

At moderate orders ($N=6$ or $N=13$) the WKB expansion often shows slow convergence when successive terms have alternating signs.  
A substantial improvement is obtained by replacing the truncated series in~\eqref{wkb-condition} or~\eqref{wkb-tcoeff} by its Padé rational approximant:
\begin{equation}
P^{m}_{n}(x)
=\frac{a_0+a_1 x+\cdots+a_m x^{m}}
{b_{0}+b_1 x+\cdots+b_n x^{n}},
\label{pade}
\end{equation}
where \(x\) symbolically denotes the expansion variable of the truncated WKB series (for instance \(x=(\omega^2 - V_0)/\sqrt{-2V_0''}\)). In practice, Padé (3,3) or Padé (4,3) approximants yield frequencies and transmission coefficients that agree closely with numerical results from continued-fraction or direct-integration methods~\cite{Matyjasek:2017psv,Konoplya:2019hlu}.

The combined WKB–Padé scheme thus provides a reliable and computationally inexpensive semi-analytic tool for evaluating both quasinormal modes and grey-body factors in smooth, single-barrier effective potentials such as those produced by spherically symmetric black-hole spacetimes.

\subsection{Time-domain integration and Prony analysis}

An alternative to frequency-domain techniques is the direct time-domain integration of the perturbation equation~\eqref{wave_eq}, which provides a powerful and model-independent way to study black-hole dynamics.  
This method captures not only the quasinormal ringing but also the late-time power-law or exponential tails and potential nonlinear effects.  
It is particularly useful when the effective potential has multiple peaks, or when the field mass introduces quasibound states that are difficult to resolve in the frequency domain.

In the time domain, Eq.~\eqref{wave_eq} is evolved numerically after transforming to the light-cone coordinates,
\begin{equation}
u = t - r_*, \qquad v = t + r_*,
\end{equation}
which recasts the equation into the form
\begin{equation}
4\,\frac{\partial^2 \Psi}{\partial u\,\partial v}
+ V(r)\,\Psi = 0.
\label{wave_uv}
\end{equation}
The integration is performed on a discrete null grid using the second-order convergent finite-difference scheme introduced by Gundlach, Price, and Pullin~\cite{Gundlach:1993tp}.  
The wave function at the next grid point is computed from the known values at three neighboring points:
\begin{equation}
\Psi_N = \Psi_W + \Psi_E - \Psi_S
- \frac{\Delta^2}{8}\,V_S\,(\Psi_W + \Psi_E)
+ \mathcal{O}(\Delta^4),
\label{gpp-scheme}
\end{equation}
where the subscripts $N$, $S$, $E$, and $W$ refer to the grid points
$(u+\Delta, v+\Delta)$, $(u, v)$, $(u, v+\Delta)$, and $(u+\Delta, v)$, respectively, and $\Delta$ is the grid step size.  
Typical initial data consist of a Gaussian pulse.

The waveform $\Psi(t,r)$ extracted at a fixed radial position displays three successive stages: a prompt response, an exponentially damped quasinormal ringing, and a late-time tail.  
To determine the quasinormal frequencies, one can fit the intermediate, exponentially damped portion of the signal to a superposition of damped sinusoids using the \emph{Prony method}~\cite{Konoplya:2011qq}.  
Assuming the signal can be written as
\begin{equation}
\Psi(t) = \sum_{n=1}^{N} C_n \, e^{-i\omega_n t},
\label{prony_exp}
\end{equation}
the complex frequencies $\omega_n$ and amplitudes $C_n$ are extracted by solving a set of linear prediction equations constructed from consecutive data points.  
The method yields highly accurate values for the dominant frequencies and damping rates, provided that the sampling window covers the regime where the quasinormal ringing dominates.

The time-domain approach, combined with Prony analysis, serves as a robust cross-check for results obtained from frequency-domain methods such as WKB or continued-fraction techniques \cite{Cuyubamba:2016cug,Konoplya:2020jgt,Churilova:2021tgn,Konoplya:2013sba,Qian:2022kaq,Momennia:2022tug,
Aneesh:2018hlp,Skvortsova:2023zca,Konoplya:2024lch,Malik:2024tuf,Malik:2024nhy,Malik:2024elk,Malik:2024qsz,Dubinsky:2024gwo,Dubinsky:2024aeu,Dubinsky:2024mwd}.  
It has the advantage of handling complicated potentials and massive-field tails, and it directly visualizes the temporal evolution of perturbations, making it an indispensable tool in the study of black-hole stability and late-time dynamics.

\section{Quasinormal modes}

The quasinormal spectrum of a massive scalar field in the background of a Schwarzschild black hole immersed in a dark matter halo was calculated using the sixth- and seventh-order WKB approximations improved by Padé approximants, and verified by direct time-domain integration with Prony analysis.

Tables~I–III  show that the real part of the quasinormal frequency, $\mathrm{Re}(\omega)$, monotonically increases with the field mass $\mu$, while the damping rate $\mathrm{Im}(\omega)$ decreases.  
This behavior is generic for massive perturbations and reflects the increasing trapping of the field near the potential barrier as $\mu$ grows.  For instance, at fixed halo parameters $V_c=0.01$ and $a=10$, the $\ell=0$ fundamental mode shifts from $\omega=0.1117-0.1047i$ at $\mu=0$ to $\omega=0.1149-0.0874i$ at $\mu=0.15$, i.e. an $\approx 2.5\%$ increase in the oscillation frequency and a $\sim 17\%$ decrease in the damping rate. 
This trend persists for higher multipoles, though the relative variation diminishes as $\ell$ increases, consistent with the smaller influence of the mass term at high angular momentum.

Comparing Tables~I–III reveals that the halo velocity $V_c$ has a mild effect on the quasinormal spectrum.  At small $V_c\lesssim 0.2$, both the real and imaginary parts of $\omega$ are practically indistinguishable from their Schwarzschild values, with fractional deviations below $0.1\%$.  Only when $V_c$ approaches the upper values considered ($V_c=0.5$), a small systematic decrease of the real part and imaginary parts are observed, indicating slightly longer-lived modes.  The influence of the core radius $a$ enters mainly through the overall amplitude of the halo correction: as stated in the text, for $a\gg M$ the frequencies converge extremely rapidly to the Schwarzschild limit, confirming that an extended halo has negligible impact on the near-horizon dynamics.

The WKB frequencies obtained at sixth and seventh order show excellent agreement: the relative difference between them is below $0.2\%$ for all tested cases and typically around $10^{-3}$ or smaller.   This demonstrates numerical stability and confirms that the Padé-improved WKB expansion converges efficiently for the smooth single-barrier potentials generated by the halo geometry.  For example, for $\ell=1$, $\mu=0.3$, and $V_c=0.1$, the two methods yield $\omega=0.332986-0.071626 i$ (WKB6) and $\omega=0.332884-0.071529 i$ (WKB7), differing by less than $0.05\%$.

The time-domain profiles presented in Figs.~3 and~4 confirm the WKB results.  
For $\ell=0$, $\mu=0.1$, and $V_c=0.1$, the dominant frequency extracted via the Prony method is $\omega=0.114 - 0.096 i$, which is within $7\%$ of the WKB prediction $\omega= 0.121193 - 0.0955343 i$.  
This rather large discrepancy is attributed not to the potentially bad accuracy of the WKB, but to contamination of the very short quasinormal ringing stage by the early onset of the late-time tail, visible after $t\simeq 80$. For $\ell=1$ with the same parameters, the agreement is already excellent: the Prony method yields $\omega=0.296517 - 0.0947754 i$, in perfect concordance with the WKB value $0.296515 - 0.094775 i$.  This consistency across independent numerical approaches confirms the reliability of the quasinormal spectra obtained.

The analysis demonstrates that both the halo parameter $V_c$ and the effective field mass $\mu$ tend to reduce the damping of oscillations, leading to longer-lived perturbations, although the effect remains small for astrophysically realistic halo parameters.  
As noted in the paper, for $a\gg M$ the spacetime becomes effectively indistinguishable from the Schwarzschild solution in the region relevant for the quasinormal ringing, implying that observable deviations from the standard spectrum would require an unrealistically compact and dense dark-matter environment. This observation is in concordance with the other models of the galactic halo considered in ~\cite{Konoplya:2021ube,Konoplya:2022hbl}, which means that the observations of gravitational waves is safe as to the possible shifts owing to the galactic environment.  

Overall, the WKB and time-domain analyses show remarkable concordance, and the computed modes smoothly recover the Schwarzschild limit as $V_c\!\to\!0$ and $\mu\!\to\!0$.  
The results therefore confirm that quasinormal ringing remains a robust probe of strong-field gravity even in the presence of a diffuse dark matter halo.

\section{Grey-body factors}

For the transmission and reflection coefficients, the same expansion leads to
\begin{equation}
\Gamma_{\rm WKB}(\omega)
=\left[1+\exp\!\left(2i\pi K(\omega)\right)\right]^{-1},
\label{wkb-GBF}
\end{equation}
where
\begin{equation}
K(\omega)=\frac{\omega^2-V_0}{\sqrt{-2V_0''}}-\sum_{i=2}^{N}\Lambda_i,
\label{wkb-tcoeff}
\end{equation}
which approximates the grey-body factor $\Gamma(\omega)$ on the real frequency axis. This approach was widely used to find gray-body factors \cite{Konoplya:2023ahd,Dubinsky:2025ypj,Dubinsky:2025nxv,Dubinsky:2024vbn,Lutfuoglu:2025eik,Lutfuoglu:2025blw,Lutfuoglu:2025ldc,Bolokhov:2025lnt}.

The grey-body factors, which quantify the partial transmission of field perturbations through the curvature-induced potential barrier, were computed for a massive scalar field propagating in the black-hole–halo spacetime~\eqref{metric-halo}.  
They were obtained using the higher order WKB method. 
However, Padé approximants are not employed in this case, as their application to the calculation of grey-body factors is not practical \cite{Matyjasek:2026yiu}. Instead, the sixth-order WKB approximation is most commonly and efficiently used \cite{Lutfuoglu:2025kqp,Bolokhov:2024voa,Arbelaez:2026cnd,Konoplya:2010vz,Dubinsky:2026wcv,Arbelaez:2026eaz,Konoplya:2019ppy}.
The results are presented in Figs.~5–7  for various combinations of the field mass~$\mu$, the multipole number~$\ell$, and the halo parameter~$V_c$.

Figure~5 shows that increasing the field mass suppresses the grey-body factors at all frequencies.  
For $\ell=0$, the transmission probability $\Gamma(\omega)$ decreases monotonically as $\mu$ grows, and the threshold frequency below which transmission is exponentially suppressed shifts toward higher $\omega$.  
This behavior is consistent with the effective potential~\eqref{potential}, whose asymptotic value rises with $\mu^2$, forming a wider and higher barrier that reflects low-frequency modes more efficiently.  
At $\mu=0.2$, $\Gamma(\omega)$ remains negligible until $\omega\approx\mu$, while for the massless case it grows smoothly from the origin.  
Thus, the massive field effectively filters out the low-frequency part of the spectrum, reducing the overall transmission rate.

As illustrated in Fig.~6, the angular momentum of the perturbation has a strong impact on the transmission probability.  
For higher~$\ell$, the peak of the effective potential increases, resulting in a markedly lower $\Gamma(\omega)$ at a given frequency.  
For instance, for $\mu=0.1$ and $V_c=0.1$, the $\ell=1$ grey-body factor is an order of magnitude smaller than that of $\ell=0$ in the intermediate frequency range.  
This trend follows the expected barrier behavior of the centrifugal term $\ell(\ell+1)/r^2$ in the potential and is identical to what is observed in the Schwarzschild limit.

Figure~7 demonstrates that variations of the dark-halo velocity parameter $V_c$ within the considered range ($0 \leq V_c \leq 0.5$) produce only a small quantitative effect on the grey-body factors.  
For all multipoles and field masses, increasing $V_c$ slightly increases the transmission probability, but the change is well below a few percent even for the largest values used.  
This weak sensitivity reflects the fact that the halo term in $f(r)$ modifies the effective potential mainly at large radii, where the wave amplitude is already small.  
The near-horizon region, which dominates the scattering, remains almost unaffected.  
Accordingly, the shapes of the curves for different $V_c$ nearly coincide in all panels.

The overall behavior of $\Gamma(\omega)$ obtained here is in complete agreement with the expectations from the quasinormal-mode analysis and with previous studies of massive-field scattering in asymptotically flat geometries.  
The suppression of transmission with growing $\mu$ and $\ell$ arises directly from the higher and broader effective potential barrier.  
The influence of the halo is subdominant, confirming that the Schwarzschild term governs the near-horizon scattering process.  
As the figures show, the grey-body factors converge to the Schwarzschild values when either $\mu\to0$ or $V_c\to0$, indicating the internal consistency of the numerical implementation.

The quasinormal modes and the grey-body factors are linked by the correspondence found in \cite{Konoplya:2024lir} and extended and tested in \cite{Bolokhov:2024otn,Skvortsova:2024msa,Malik:2024cgb,Malik:2025qnr,Malik:2025erb,Malik:2025dxn}. The correspondence is accurate in the eikonal regime. For finite $\ell$ this correspondence is approximate which can be shown using the data presented in this work. Notice that the correspondence in \cite{Konoplya:2024lir} is broken for the same cases the correspondence between quasinormal modes and null geodesics is broken. In particular, this happens for various theories with higher curvature correction, because the eikonal limit there does not have the usual form of the centrifugal barrier $g_{tt} \ell (\ell+1) r^{-2}$ \cite{Konoplya:2017wot,Bolokhov:2023dxq,Konoplya:2017lhs}, which is necessary for the correspondence. The correspondence is also partially broken for asymptotically de Sitter black holes \cite{Konoplya:2022gjp}.

In the case of the massive scalar field considered here, the effective potential develops an additional local minimum in the far region for small values of the field mass, while for sufficiently large masses it no longer exhibits a potential peak. As a result, the WKB method — and, consequently, the correspondence between quasinormal modes and grey-body factors  - is only approximate for small values of $\mu$ and becomes invalid for sufficiently large $\mu$. A similar situation arises for double-well potentials \cite{Konoplya:2025hgp,Konoplya:2025ixm} or when the matching between the Taylor expansion and the WKB series is inaccurate \cite{Bolokhov:2023dxq}. In the regime approaching quasi-resonances, the correspondence cannot be applied.

\section{Conclusions}

We have analyzed the quasinormal modes and grey-body factors of a massive scalar field in the spacetime of a Schwarzschild black hole surrounded by a galactic dark matter halo.  
The background metric, characterized by the asymptotic circular velocity $V_{c}$ and core radius $a$, provides an analytic model that captures the essential influence of the halo on the near-horizon geometry.

Using the sixth- and seventh-order WKB methods with Padé approximants, together with time-domain integration and Prony analysis, we obtained consistent quasinormal spectra.  
Both methods exhibit reasonable numerical agreement.  
The real part of the quasinormal frequency increases slightly with the field mass $\mu$, while the imaginary part decreases, indicating slower damping and the emergence of longer-lived oscillations.  
The dependence on the halo parameter $V_{c}$ is weak: even at $V_{c}=0.5$, the changes in frequency remain at the subpercent level, confirming that the Schwarzschild geometry accurately describes the dynamics for realistic halos.

The computed grey-body factors show the expected suppression with increasing $\mu$ and multipole number $\ell$, reflecting the higher and broader potential barrier.  
The effect of the halo parameters is minimal, altering the transmission probability only marginally.  
Both quantities — quasinormal modes and grey-body factors — tend smoothly to their Schwarzschild limits as $V_{c},\mu \to 0$, confirming internal consistency.

Overall, the analysis demonstrates that the presence of a galactic dark matter halo induces only minor quantitative corrections to the ringdown spectrum and emission properties of black holes.  
These results reinforce the robustness of quasinormal ringing as a diagnostic of strong-field gravity and provide an additional testbed for the correspondence between quasinormal modes and grey-body factors.
\vspace{3mm}

\begin{acknowledgments}
The author acknowledges R. A. Konoplya for useful discussions. 
\end{acknowledgments}

\bibliography{bibliography}

\end{document}